\documentclass{cs20proc}

\usepackage{color}

\editors{S. J. Wolk, A. K. Dupree, H. M. G\"unther}

\publisher{Zenodo}

\conference{The 20.5th Cambridge Workshop on Cool Stars, Stellar Systems, and the Sun}

\conferencedate{2021}

\title{Planet-induced radio emission from the coronae of M dwarfs}

\author{Robert D. Kavanagh,$^{1}$ Aline A. Vidotto,$^{1}$ Baptiste Klein,$^{2}$ Moira M. Jardine,$^{3}$ Jean-Fran\c{c}ois Donati,$^{4}$ D\'{u}alta \'{O} Fionnag\'{a}in$^{5}$}

\affiliation{$^{1}$School of Physics, Trinity College Dublin, The University of Dublin, Dublin 2, Ireland \\
$^{2}$Sub-department of Astrophysics, Department of Physics, University of Oxford, Oxford OX1 3RH, UK \\
$^{3}$SUPA, School of Physics and Astronomy, University of St Andrews, St Andrews KY16 9SS, UK \\
$^{4}$Universit\'e de Toulouse, CNRS, IRAP, 14 av. Belin, 31400 Toulouse, France \\
$^{5}$Centre for Astronomy, National University of Ireland, Galway, Ireland}

\shorttitle{Planet-induced radio emission from M dwarfs}

\shortauthors{Kavanagh et al.}

\abs{There have recently been detections of radio emission from low-mass stars, some of which are indicative of star-planet interactions. Motivated by these exciting new results, here we present stellar wind models for the active planet-hosting M dwarf AU Mic. Our models incorporate the large-scale photospheric magnetic field map of the star, reconstructed using the Zeeman-Doppler Imaging method. We use our models to assess if planet-induced radio emission could be generated in the corona of AU Mic, through a mechanism analogous to the sub-Alfvénic Jupiter-Io interaction. In the case that AU Mic has a mass-loss rate of 27 times that of the Sun, we find that both planets b and c in the system can induce radio emission from 10 MHz -- 3 GHz in the corona of the host star for the majority of their orbits, with peak flux densities of 10 mJy. Our predicted emission bears a striking similarity to that recently reported from GJ 1151 by \citet{vedantham20}, which is indicative of being induced by a planet. Detection of such radio emission would allow us to place an upper limit on the mass-loss rate of the star.}

\begin{document}

\maketitle



\section{Introduction}

Many theoretical works have aimed to identify potential targets for the detection of low frequency exoplanetary radio emission \citep{griessmeier07, saur13, vidotto15, vidotto17, turnpenney18, kavanagh19, kavanagh20}. One such model used in these works is analogous to the sub-Alfv\'enic interaction between Jupiter and its moon Io \citep{neubauer80, zarka98, saur04, zarka07, griessmeier07}, with the host star and planet taking the roles of Jupiter and Io respectively. If the planet orbits with a sub-Alfv\'enic velocity relative to the wind of its host star, it can generate Alfv\'en waves that travel back towards the star \citep{ip04, mcivor06, lanza12, turnpenney18, strugarek19, vedantham20}. A fraction of the wave energy produced in this interaction is expected to dissipate, producing radio emission via the electron cyclotron maser instability (ECMI) in the corona of the host star \citep{turnpenney18}.

Thanks to the increasing sensitivity of radio telescopes such as LOFAR, M~dwarfs are beginning to light up the radio sky at low frequencies (Callingham et al., submitted). One such system which was recently detected to be a source of emission from 120 -- 160~MHz by \citet{vedantham20} is the quiescent M~dwarf GJ~1151. The authors illustrated that the observed emission is consistent with ECMI from the star induced by an Earth-sized planet orbiting in the sub-Alfv\'enic regime with a period of 1 -- 5 days. Prior to this detection, there had been no evidence to suggest GJ~1151 is host to a planet. There has been some discussion in the literature recently about the existence of such a planet. \citet{mahadevan21} have suggested that a planet orbits the star in a 2-day orbit, whereas \citet{perger21} have ruled this out, placing a mass upper limit of 1.2 Earth masses on a planet in a 5-day orbit. Follow-up observations of the system will be needed to further assess if the radio emission is of a planet-induced origin.

AU Microscopii (AU Mic) is a young M dwarf that shows potential for the detection of planet-induced radio emission. It lies just under 10~pc away from Earth, and has recently been discovered to host two Neptune-sized close-in planets \citep{plavchan20, martioli21}. While planets b and c orbiting AU Mic are not likely to be habitable, their proximity to the host star makes them ideal candidates for being in sub-Alfv\'enic orbits and inducing radio emission in the corona of the host star. Here we present stellar wind modelling of AU Mic, which we use to assess whether planet-induced radio emission could be generated in the system. The full results of this work are published in \citet{kavanagh21}.



\section{Stellar wind environment of AU Mic}

\begin{figure*}[!t]
\includegraphics[width = 0.495 \textwidth]{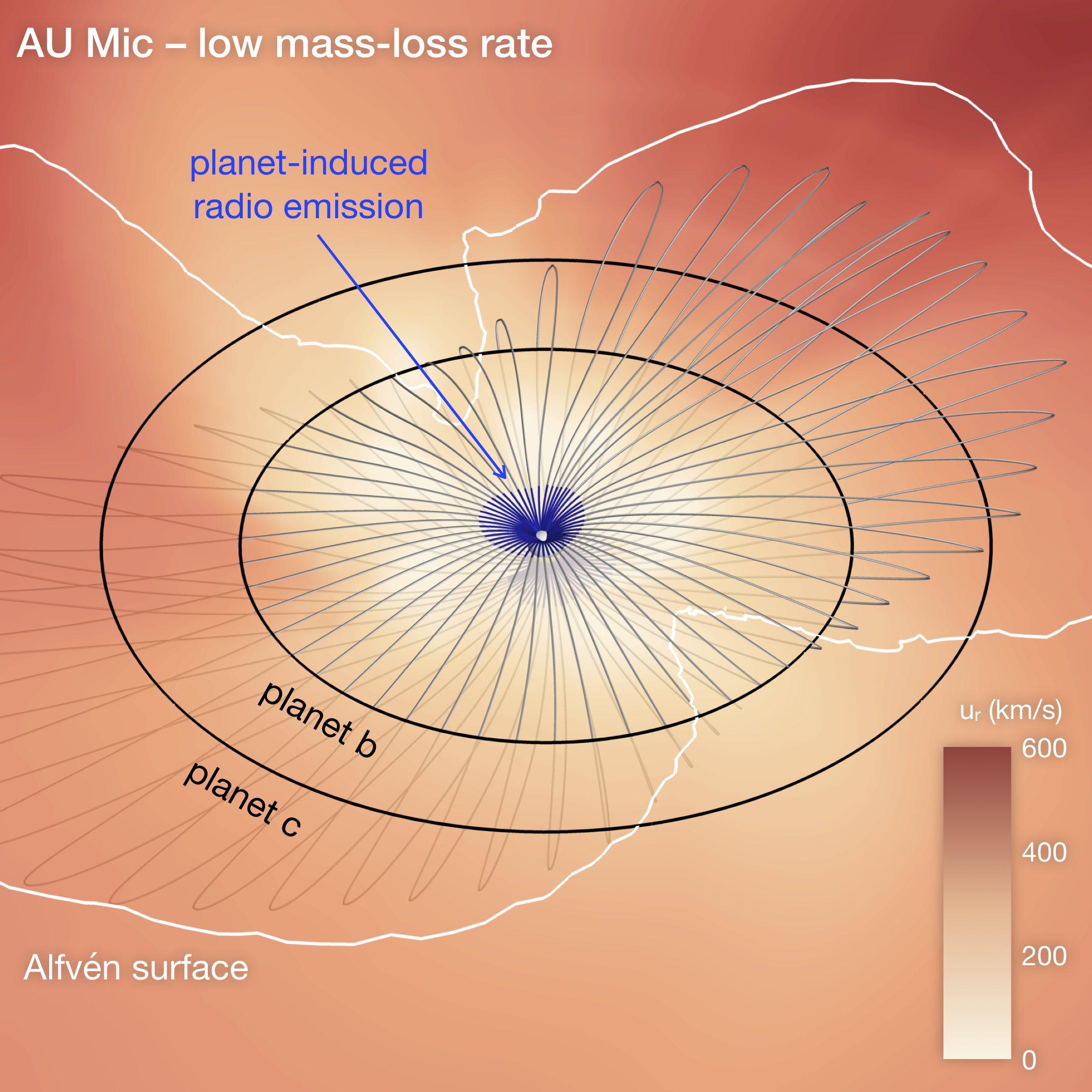}
\includegraphics[width = 0.495 \textwidth]{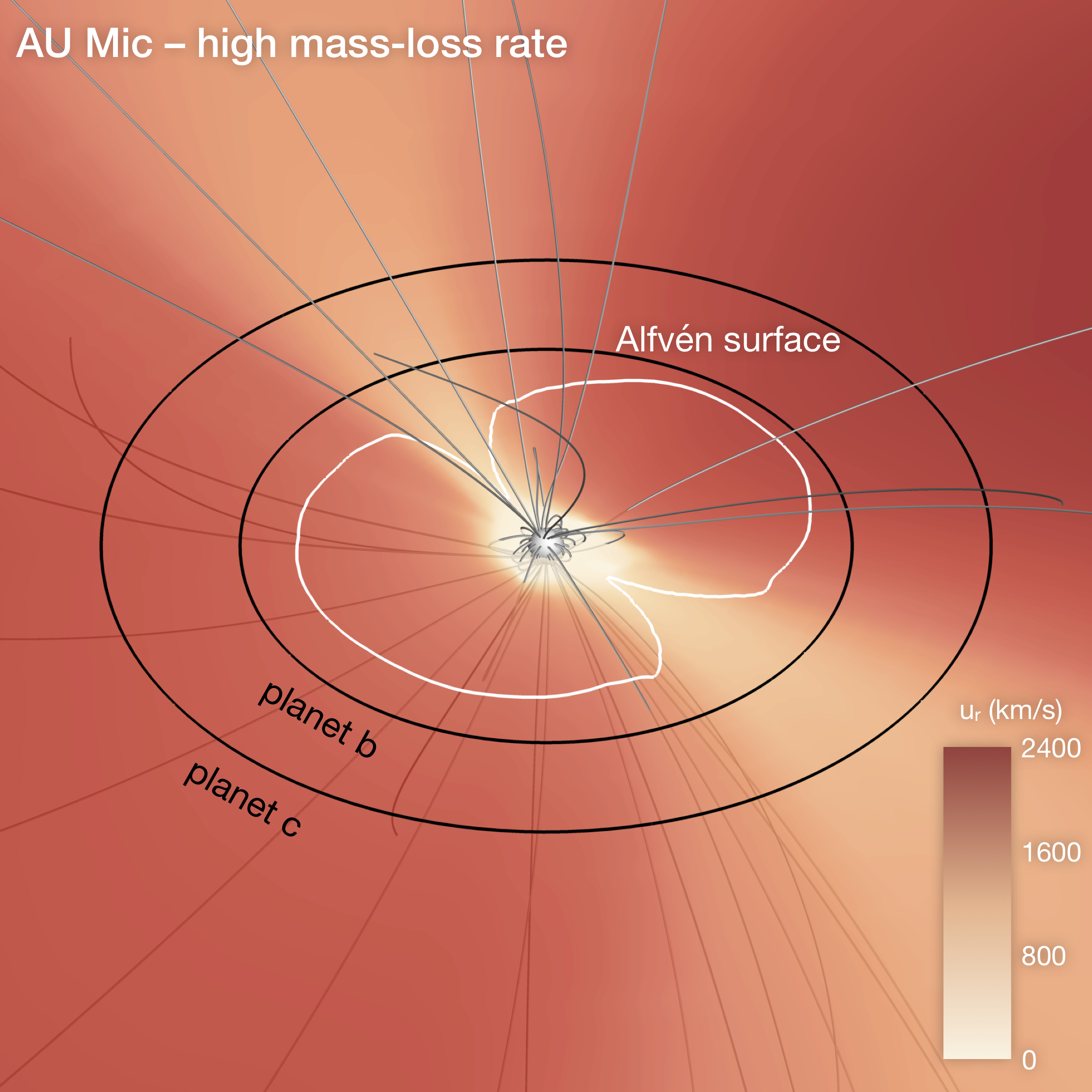}
\caption{\textit{Left}: Low $\dot{M}$ model of the stellar wind of AU Mic. The orbits of planets b and c are shown as black circles, and the white line corresponds to the Alfv\'en surface. Inside this surface, the planets are said to orbit sub-Alfv\'enically (see Equation~\ref{eq:alfven velocity}), and can induce radio emission in the star's corona. The contour in the orbital plane is coloured by the wind radial velocity ($u_r$). The grey lines show the stellar magnetic field lines that connect to the orbit of planet b. Each of these lines is a closed loop, and connects back to the star in both the Northern and Southern hemisphere. The blue shaded region of each line illustrates where planet b can induce radio emission via ECMI (see Equation~\ref{eq:plasma frequency}). Note that planet c can also induce emission, but for clarity we omit these field lines. \textit{Right}: High $\dot{M}$ model for AU Mic. Both planets b and c orbit in the super-Alfv\'enic regime in this scenario. Note that the magnetic field lines shown here do not connect to the orbit of either planet.}
\label{fig:wind models}
\end{figure*}

In order to determine if the orbiting planets around AU Mic can induce radio emission in its corona, we first perform 3D magnetohydrodynamic (MHD) stellar wind simulations for AU Mic using the Alfv\'en wave-driven AWSoM model \citep{vanderholst14} implemented in the BATS-R-US code \citep{powell99}. We impose the large-scale surface magnetic field map of the star from \citet{klein21} as a boundary condition in our models. The stellar wind mass-loss rate of AU Mic is relatively unconstrained. Models of interactions between the stellar wind and debris disk in the system estimate a mass-loss rate from 10~$\dot{M}_{\odot}$ \citep{plavchan09} up to 1000~$\dot{M}_{\odot}$ \citep{chiang17}. In order to explore these two scenarios of low and high mass-loss rates, we vary the flux of Alfv\'en waves that drive the stellar wind outflow \citep[for full details see][]{kavanagh21}.

Using our two inputs of Alfv\'en wave fluxes, we obtain mass-loss rates for AU Mic of 27 and 590~$\dot{M}_\odot$ ($\dot{M}_\odot = 2\times10^{-14}~M_\odot$~yr$^{-1}$). We refer to these as the `Low $\dot{M}$' and `High $\dot{M}$' models respectively. Figure~\ref{fig:wind models} show 3D views of both models. We see in the case of the Low $\dot{M}$ model, both planets are in sub-Alfv\'enic orbits for the majority of the time, satisfying the condition needed for the planet to induce radio emission in the corona of the host star.



\section{Modelling planet-induced radio emission}

\begin{figure}[!t]
\includegraphics[width = \columnwidth]{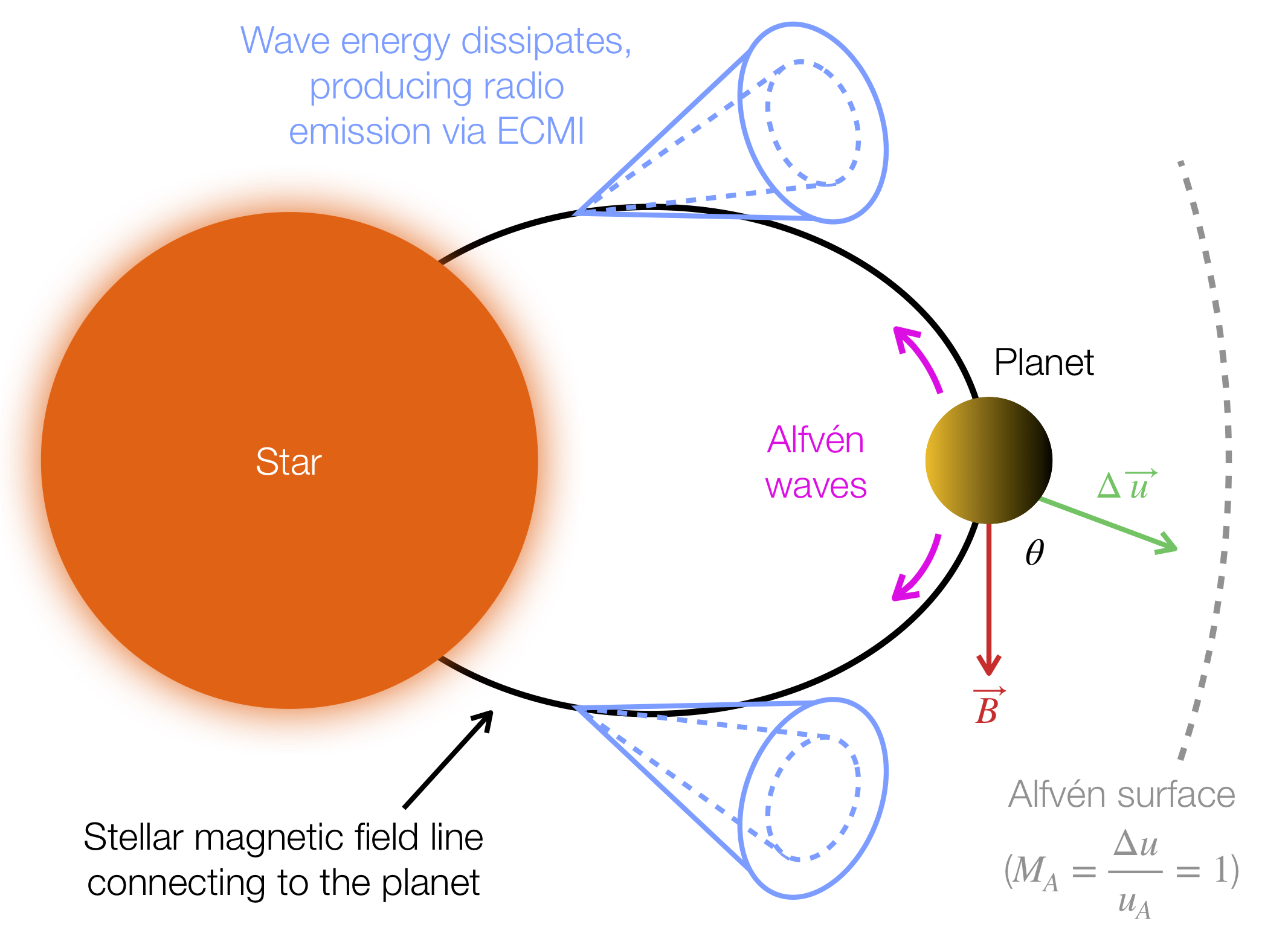}
\caption{Sketch of our planet-induced radio emission model. A planet orbiting inside the Alfv\'en surface of the wind of its host star can perturb the stellar magnetic field, producing Alfv\'en waves. These waves then travel back towards the host star along the magnetic field line connecting the star and planet, producing radio emission via the electron cyclotron maser instability (ECMI). The emission is beamed in a cone. Close to the star, the planet orbits through the closed-field region of the star's magnetic field. Therefore, emission can be generated in both the Northern and Southern hemispheres of the star.}
\label{fig:sketch}
\end{figure}

We now describe our model we use to calculate the expected flux density and frequency of the emission induced by the two planets in the corona of AU Mic. We use the stellar wind properties at the two planetary orbits from our Low $\dot{M}$ model in our calculations. A sketch of our model is shown in Figure~\ref{fig:sketch}.

When the planet is in a sub-Alfv\'enic orbit (i.e., inside the Alfv\'en surface), the relative velocity between the wind and orbit $\Delta u$ is less than the Alfv\'en velocity:
\begin{equation}
\Delta u < u_A = \frac{B}{\sqrt{4\pi\rho}}.
\label{eq:alfven velocity}
\end{equation}
Here, $B$ and $\rho$ are the magnetic field strength and density of the stellar wind at the planet's orbit. Inside the Alfv\'en surface, the planet perturbs the magnetic field of the star, producing Alfv\'en waves which travel back towards the star along the stellar magnetic field line connecting the star and planet. Close to the star, the planet orbits in the closed field region of the stellar magnetic field (see the left panel of Figure~\ref{fig:wind models} and Figure~\ref{fig:sketch}). As a result, emission can be generated in both the Northern and Southern hemisphere of the star's corona.

The wave power produced in this interaction is \citep{saur13}:
\begin{equation}
P = \pi^{1/2} R^2 B \rho^{1/2} \Delta u^2 \sin^2 \theta.
\end{equation}
Here, $R$ is the radius of the obstacle perturbing the stellar magnetic field. For an unmagnetised planet, we take this as the planetary radius. $\theta$ is the angle between the vectors $\vec{B}$ and $\Delta\vec{u}$. A fraction of this energy \citep[$\sim1\%$,][]{turnpenney18} is converted into radio emission, producing a flux density observed at a distance $d$ from the system of
\begin{equation}
F_\nu = \frac{0.01 \times P}{\Omega d^2 \nu}.
\end{equation}
The emission is beamed in a cone with a solid angle of $\Omega = 1.6$~sr \citep{zarka04}. It is generated at the cyclotron frequency $\nu = 2.8~B$~MHz, where $B$ is the magnetic field strength in gauss (G) at the emitting point along the field line connecting the star and planet. Note that in order for the emission to be generated, the cyclotron frequency must exceed the plasma frequency
\begin{equation}
\nu > \nu_\textrm{\small p} = 9\times10^{-3}\sqrt{n_e}~\textrm{MHz},
\label{eq:plasma frequency}
\end{equation}
where $n_e$ is the electron number density. These regions of the stellar magnetic field lines connecting to the orbit of planet b are shown in blue in Figure~\ref{fig:wind models}.

\begin{figure}
\includegraphics[width = \columnwidth]{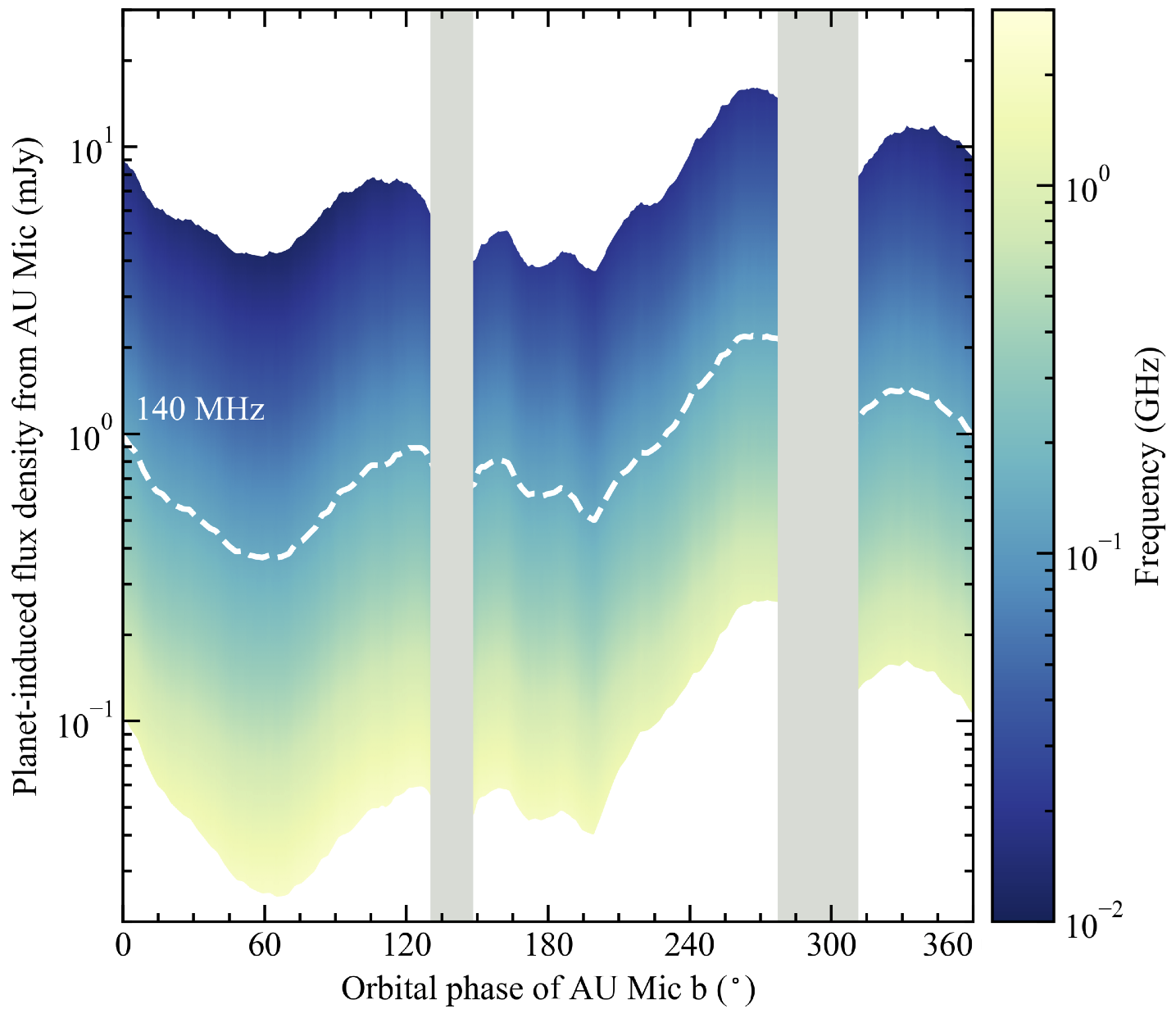}
\caption{Radio spectrum of AU Mic induced by planet b in the Northern hemisphere of the star's corona as a function of orbital phase. The emission corresponds to the blue regions of the magnetic field lines connecting to the planet's orbit shown in the left panel of Figure~\ref{fig:wind models}. Emission generated at 140~MHz is highlighted with a white dashed line. This is the middle frequency of the observing band at which radio emission was recently detected from the M~dwarf GJ~1151 by \citet{vedantham20}, which is suspected of being induced by an orbiting planet. The grey shaded areas illustrate the region where the orbit of the planet is in the super-Alfv\'enic regime, where no emission can be induced by the planet.}
\label{fig:spectrum}
\end{figure}

In Figure~\ref{fig:spectrum} we show the flux density induced by planet b in the Northern hemisphere of AU Mic as a function of its orbital phase. The flux density is colour-coded with the emission frequency. We find that planet b can induce emission from 10~MHz -- 3~GHz, with peak flux densities of 10~mJy. Note that similar emission can also be induced in the Southern hemisphere \citep[see Figure 4 of][]{kavanagh21}. The emission induced by planet c is also in this frequency region, albeit at flux densities that are an order of magnitude lower. 

We highlight the emission at 140~MHz in Figure~\ref{fig:spectrum}, which is the the middle of the frequency range of 120 -- 160~MHz at which some M~dwarfs have recently been detected \citep[][Callingham et al., submitted]{vedantham20}. \citet{vedantham20} suggested that their observations of emission from the M dwarf GJ~1151 may be generated by a planet orbiting in the sub-Alfv\'enic regime. At 140~MHz, our results bear a strong resemblance to the observations of GJ~1151: both have flux densities of about 1~mJy which exhibit temporal variations. It would be very useful to obtain magnetic field maps and near-simultaneous radio observations of M~dwarfs similar to the AU Mic planetary system so that this scenario could be explored further. Detection of such emission allow us to constrain the mass-loss rate of the host star, which in the case of AU Mic we have shown to be possible if it has a mass-loss rate of 27~$\dot{M}_\odot$. This would also compliment observations of the planetary transits in the ultraviolet, which have been shown to vary depending on the mass-loss rate of the host star \citep{carolan20}.

\section{Conclusions}

Here we have presented the results published in \citet{kavanagh21}, wherein we performed stellar wind simulations of the active planet-hosting M dwarf AU Mic, exploring the two cases where the star has a low and high mass-loss rate. We used our models to investigate if the orbiting planets could induce radio emission in the corona of the star. In the case of a low mass-loss rate ($\dot{M} = 27~\dot{M}_\odot$), we found that both planets b and c can induce emission in AU Mic's corona from 10~MHz -- 3~GHz. For planet b, the induced flux density peaks at 10~mJy, with that induced by planet c being an order of magnitude lower. We also illustrated that our predicted emission induced by planet b at 140~MHz bears a striking resemblance to that reported for GJ~1151 by \citet{vedantham20}, which is suspected of being induced by an orbiting planet. In \citet{kavanagh21}, we also investigated the potential for planet-induced radio emission to be generated in the corona of the active M dwarf Prox Cen. However, as we were not able to find a scenario where this could occur, we have omitted these results here for brevity. This work has illustrated how detection of planet-induced radio emission could be used to constrain the mass-loss rate of the host star.


\section*{Acknowledgments}

We thank the SOC for giving us the opportunity to present this work. RDK acknowledges funding received from the Irish Research Council (IRC) through the Government of Ireland Postgraduate Scholarship Programme. AAV, BK, and JFD acknowledge funding from the European Research Council (ERC) under the European Union's Horizon 2020 research and innovation programme (grant agreement No 817540, ASTROFLOW, 865624, GPRV, 740651, NewWorlds). MMJ acknowledges support from STFC consolidated grant number ST/R000824/1. D\'{O}F acknowledges funding from the IRC Government of Ireland Postdoctoral Fellowship Programme. We acknowledge the Irish Centre for High-End Computing (ICHEC) for providing the computational facilities used to perform the simulations published in this work.


\bibliographystyle{cs20proc}
\bibliography{bibliography.bib}

\end{document}